\documentclass[epj]{svjour}
\usepackage{graphicx}
\usepackage{amsmath,amssymb}
\graphicspath{{aFigures/}}
\usepackage{epstopdf}
\begin{document}
\title{Dry microfoams:\\Formation and flow in a confined channel}
\author{Jan-Paul Raven  \and Philippe Marmottant \and Fran\c cois Graner
\thanks{\emph{E-mail:} jpraven@spectro.ujf-grenoble.fr}%
}                     
%
%
\institute{Laboratoire de Spectrom\'etrie Physique, CNRS - UMR 5588,
Universit\'e Grenoble I,  B.P. 87,
F-38402 St Martin d'H\`eres Cedex, France}

\date{\today}
%
\abstract{
We present an experimental investigation of the agglomeration of 
microbubbles into a 2D microfoam and its flow in a rectangular microchannel. Using a 
flow-focusing method, we produce the foam \textit{in situ} on a 
microfluidic chip for a large range of liquid fractions, down to a few percent in liquid. 
We can monitor the transition from separated bubbles to the desired 
microfoam, in which bubbles are closely packed and separated by thin films. 
We find that bubble formation frequency is limited by the liquid flow rate, whatever the gas pressure.  The formation  frequency creates a modulation of the foam flow, rapidly damped along the channel. The average foam flow rate depends non-linearly on the applied gas pressure, 
displaying a threshold pressure due to capillarity. 
Strong discontinuities in the  flow rate appear when the number of bubbles in the channel width changes, reflecting  the discrete nature of the foam topology. 
 We also produce an ultra flat foam, reducing the channel height from 250 $\mu$m to  8 $\mu$m, resulting in a height to diameter ration of 0.02; we notice a marked change in bubble shape during the flow.
\PACS{
      {47.60.+i}{Flows in ducts, channels, nozzles, and conduits} \and
      {83.50.Ha}{Flow in channels} \and
      {83.80.Iz}{Emulsions and foams}
     } 
} 
\maketitle

\section{\label{sec:Intro} Introduction}
Two phase microflows like microemulsions, microbubbles and microdrops presently attract considerable attention  \cite{Link2004,Cubaud2004,Garstecki2004,Garstecki2005,Drenckhan2005}. To these we would like to add \emph{microfoams}. Their application in a lab-on-a-chip context provides the possibility for the efficient handling of series of gas pockets, and allows to create microchemical  reactors that are both very rapid and highly parallelized.  
Specifically,  the gas-liquid interface of  microfoams provides a transport location for amphiphilic molecules, with a hydrophilic head and a hydrophobic tail. A decrease in size increases the surface to volume ratio; hence, microfoams could be used as an efficient carrier for proteins or lipids at high concentration. 

Microfoams offer  advantages compared to foams at larger scales for the study of foam properties as microfoams are very stable and well controlled.  First, the absence of vertical drainage on the small length scales of a microfluidic system creates liquid profiles in the foam films that are constant over time and do not show an asymmetry due to gravity. Another advantage is that because of the low Reynolds numbers involved, the amount of gas produced during bubbling is very stable with  nearly  monodisperse bubble volume distribution. The amount of liquid (liquid fraction) can be reproducibly controlled as it is governed by the input parameters.

A set of basic operations using specific channel geometries necessary for manipulating series of bubbles, termed ``discrete microfluidics", has been demonstrated  at the millimeter scale \cite{Drenckhan2005} in a ``dry" foam, where the liquid content is low compared to the gas content. To down-scale these operations and adapt them for microfluidics requires producing a  microfoam and information about its flow characteristics. 

Here, we investigate the continuous production of a two phase gas-liquid flow in a flow-focusing device,  and the transitions between different regimes of bubble formation, so as to reach  microfoams, thereby extending the studies of Ref. \cite{Garstecki2004} to low liquid fractions.  In a microfluidic flow-focusing device,  a flowing gas thread  is forced, by the co-flowing surrounding liquid, into a small orifice, where the gas thread breaks up at regular time intervals \cite{Garstecki2004,Garstecki2005,Garstecki2005b}. At low liquid to total flow rate ratio this will create a microfoam. 

We would like to determine how  foam properties (for a review, see \cite{Weaire1999}) extrapolate to the micrometer range.  We investigate here microfoam formation and flow  dissipation within a microfluidic set-up: a bubble formation orifice, followed by a long channel ending with a free  exit. We finally open perspectives for the  study of structure and dissipation of ultra-flat microfoams.

\section{\label{sec:Methods} Materials and methods}

We use a flow focusing geometry: an inlet channel for the liquid, another one for the gas, both ending in an orifice ending up in a straight channel \cite{Link2004,Garstecki2004,Garstecki2005,Garstecki2005b}. 
We produce the microfluidic device by soft lithography techniques. We first create a mold in a negative photosensitive material (SU-8 2100, MicroChem) and then make imprints in a polymer (polydimethylsiloxane, PDMS, Sylgard 184) to create the actual channel. The PDMS imprint is glued to a glass cover slide using a home-built ozone cleaner. The exit channel has a height $h = 250$ $\mu$m, width $w = 700$ $\mu$m and an orifice width $w_{or} = 100$ $\mu$m. We also use a channel where, at a distance of 6 mm after the orifice, the foam  flows through a second constriction diminishing the channel width to $125$ $\mu$m. 
We produce the ultra-flat foam  in a device with a height of about $h = 8$ $\mu$m, width $w = 400$ $\mu$m and an orifice width $w_{or} = 75$ $\mu$m; we create the mold with a positive photo-resist (ma-P 100, Micro Resist Technology). 

For the continuous phase we use deionized water with 10 $\%$ commercial dishwashing detergent (Dreft, Procter $\&$ Gamble). This solution has a surface tension $\sigma = 38 \pm 1$ mN/m, as measured by the Wilhelmy balance method. 
The use of this surfactant resulted in an increased  wettability of the solution to the PDMS surfaces
\cite{Dou2002}.
Two different syringe pumps were used to push the liquid (11 Pico Plus, Harvard Apparatus, 
and KDS 100, KD Scientific) at 
flow rates $Q_l$ ranging from $4$ to $167$ $\mu$l/min, with $\pm0.5\%$ accuracy. 
The dispersed gas phase is nitrogen. It is driven at constant overpressure $P_g$ (relative to one atmosphere), ranging from $1$ to $21$ kPa, using a 
pressurized tank and a pressure-reduction valve (stability $\pm 0.15$ kPa). Pressure $P_{in}$ is measured at the entrance of the device with piezo-resistive gauge (40PC Honeywell, $\pm 0.2$ kPa accuracy). Since the exit is at atmospheric pressure $P_{out}= 1$ atm,  the overpressure  $P_g$ is  total gas pressure drop $P_{in}-P_{out}$ over the microfluidic system (orifice and channel). 

\begin{figure}[htbp]
\setlength{\unitlength}{1cm}
\begin{picture}(6.8,5.7)(0.0,0.0)
\put(0.4,5){(a)}
\put(0.0,0){\includegraphics[scale=0.666]{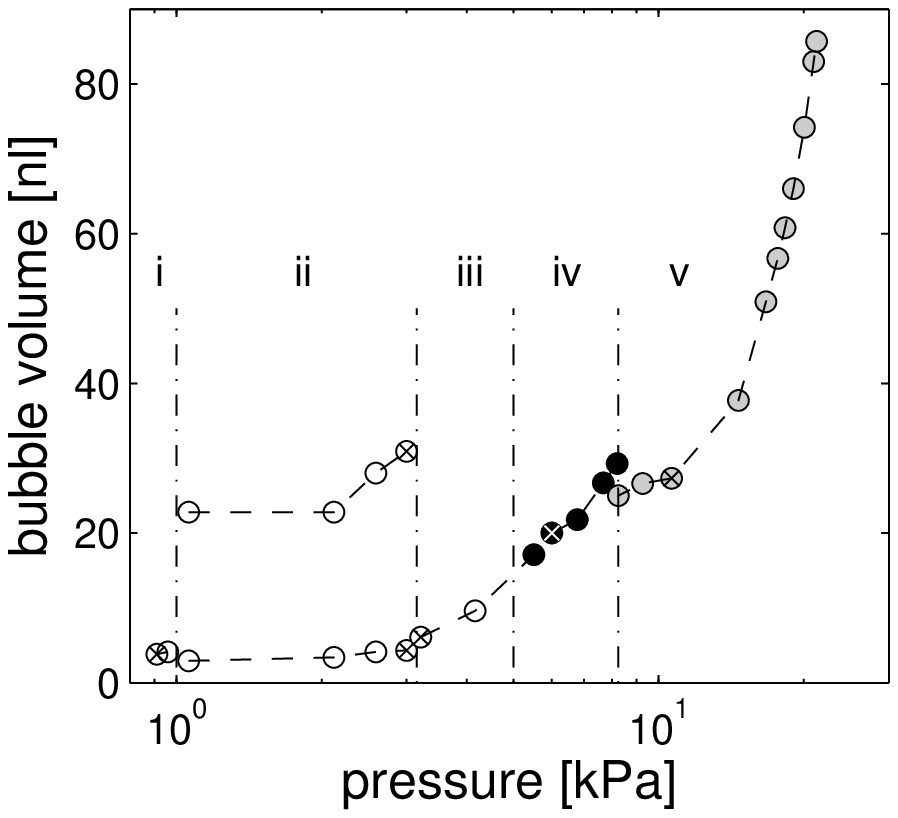}}
\end{picture}
\\
 \hspace{4mm}
 \begin{minipage}[c]{5.5cm}
\setlength{\unitlength}{1cm}
\begin{picture}(6.8,2.7)(0.0,0.0)
\put(0,2.2){(b)}
\put(0,0){\includegraphics[width=5.5cm]{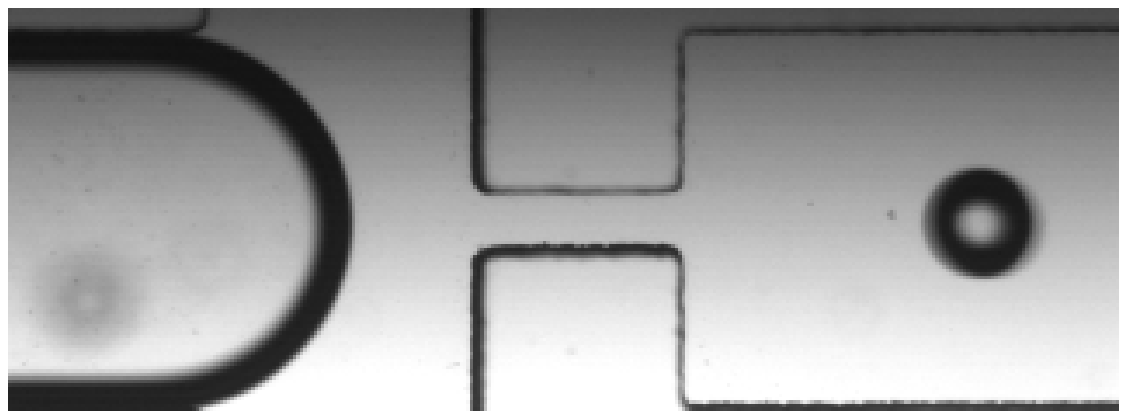}}
\put(0.5,1.3){\textsf{\textbf{gas}}}
\put(1.5,1.7){\textsf{\textbf{liquid}}}
\end{picture}
 \end{minipage}
 \hspace{0.4cm}
 \begin{minipage}[c]{2cm}
 (i) dripping 
 \end{minipage}
  \\
 \hspace{4mm}
 \begin{minipage}[c]{5.5cm}
\includegraphics[width=5.5cm]{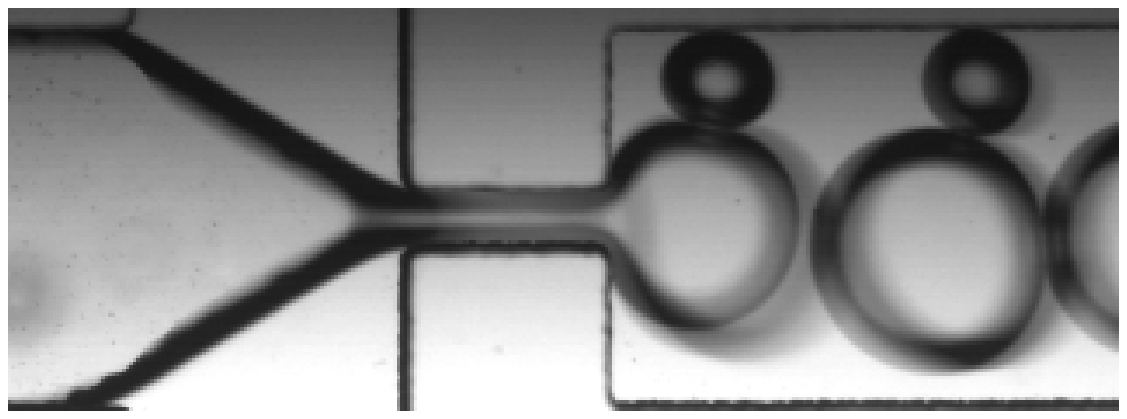} 
 \end{minipage}
  \hspace{0.4cm}
\begin{minipage}[c]{2cm}
 (ii) bidisperse \\Ê
 bubbles 
 \end{minipage}
 \\
 \hspace{4mm}
 \begin{minipage}[c]{5.5cm}
\includegraphics[width=5.5cm]{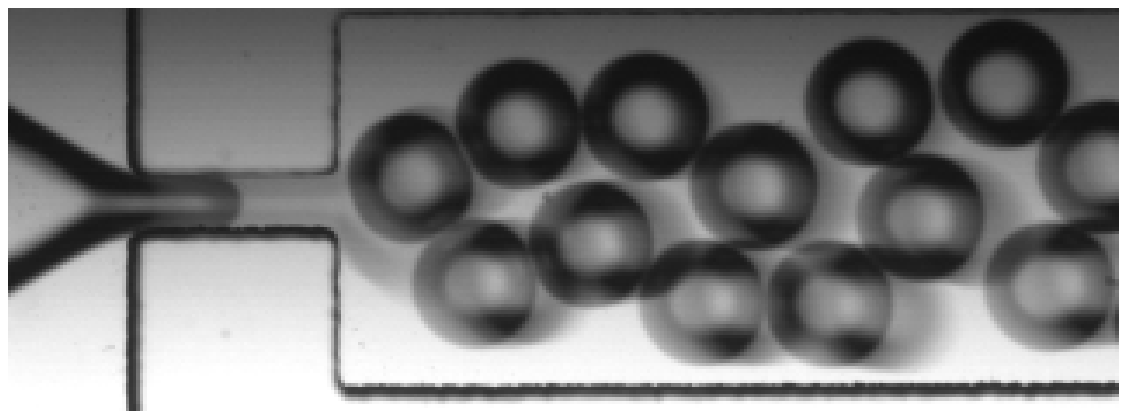} 
 \end{minipage}
  \hspace{0.4cm}
\begin{minipage}[c]{2cm}
 (iii) bubbly flow
 \end{minipage}
 \\
 \hspace{4mm}
 \begin{minipage}[c]{5.5cm}
\includegraphics[width=5.5cm]{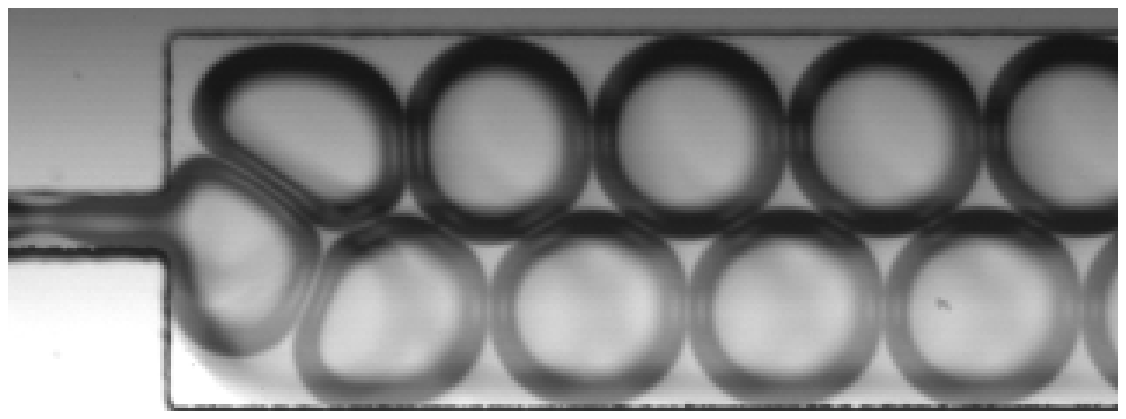} 
 \end{minipage}
  \hspace{0.4cm}
\begin{minipage}[c]{2cm}
 (iv) alternate foam
 \end{minipage}
 \\
 \hspace{4mm}
 \begin{minipage}[c]{5.5cm}
\includegraphics[width=5.5cm]{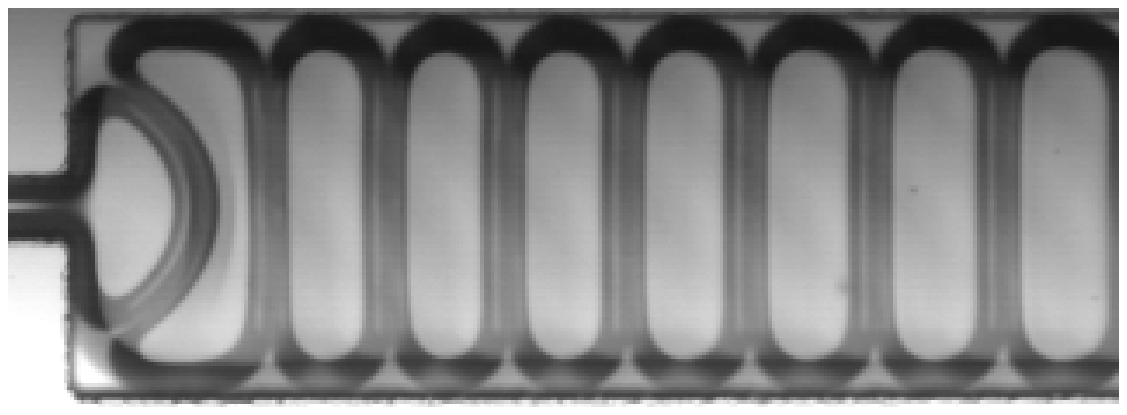} 
 \end{minipage}
  \hspace{0.4cm}
\begin{minipage}[c]{2cm}
 (v) bamboo foam 
 \end{minipage}
\caption{\label{fig:regimes}
From wet to dry microfoams. (a) Bubble volume  $V_b$ versus gas pressure 
$P_g$. Liquid flow rate is kept constant at $Q_l=167$ $\mu$l min$^{-1}$. 
Numbers identify the different regimes: (i) dripping flow; (ii) bidisperse bubbles (two symbols are plotted for each pressure); (iii) bubbly flow; (iv) alternate  foam (2 rows) with filled symbols; (v) bamboo  foam (1 row) with gray symbols.
  (b) Photographs of these regimes.
Crossed symbols in (a) correspond to pictures in (b).
}
\end{figure}

In a typical experiment we vary the gas pressure while keeping the liquid flow rate constant. In this way we scan the complete pressure range for which bubbles are formed at that liquid flow rate. Still images or movies of the resulting flow are then captured with a camera (Marlin F131B, Allied Vision 
Technologies) connected to an inverted optical microscope (IX70, 
Olympus), see Fig. \ref{fig:regimes}b for some examples. By analysing these
we can extract the following bubble quantities: formation rate (break-up frequency $f$), with a 
precision of a few percent, the bubble volume $V_b=A_b h$ by measuring 
the apparent area $A_b$ occupied by the gas in the images, the gas flow rate estimated as  $Q_g= V_b f$; and the space and time averaged gas velocity $\langle\overline{u_g} \rangle$ estimated as the distance between two consecutive bubbles multiplied by $f$. The edge of the measured area $A_b$ (measured with a precision $\pm 1\%$) is taken in the middle of the curved meniscus around the bubble, appearing black on images, systematically underestimating (with up to $10 \%$ for non-touching bubbles) the actual gas volume.

An important parameter in characterising a foam is the foam volume liquid fraction $\Phi_l$, \emph{i.e.} the proportion of the volume occupied by the liquid:
 \begin{equation}
\Phi_l=\frac{V_l} {V_g+V_l}.  
 \end{equation}
 It is estimated by image analysis with $\Phi_l \simeq 1 - {A_b}/{A}$.
Another method is to measure simultaneously the time and space averaged  bubble velocity ${\langle \overline{u_g} \rangle}$ together with the average gas flow rate; the liquid fraction then follows from $Q_g={\langle \overline{u_g} \rangle} S(1-\Phi_l)$. 

 On the other hand, the proportion of liquid \emph{injected} in the system is 
 \begin{equation}
 \alpha_l=\frac{Q_l} {Q_g+Q_l},  
 \end{equation}
with $Q = Q_g + Q_l$  the total two-phase flow. It is simply measured  as $Q_l$ is one of the control parameters. The precision in this measurement is at minimum $\pm 10 \%$ for non-touching bubbles and in general a few percent. 
The two quantities $\Phi_l$ and $\alpha_l$ are different because liquid and gas can have different velocities. They can be linked using the time and space averaged gas and liquid velocities $\langle\overline{u_g} \rangle$ and $\langle\overline{u_l} \rangle$ since $Q_l=\langle\overline{u_l} \rangle S \Phi_l$ and $Q_g=\langle\overline{u_g} \rangle S (1-\Phi_l)$ with $S$ the area of the channel cross section. This yields: 
\begin{equation}
\frac{\langle \overline{u_l} \rangle}{\langle \overline{u_g} \rangle}=\frac{\alpha_l}{1-\alpha_l}\frac{1-\Phi_l}{\Phi_l}.	
\label{eq:massconservation}
\end{equation}
The separate measurement of $\alpha_l$ and $\phi_l$  allows to calculate the ratio ${\langle \overline{u_l} \rangle}/{\langle \overline{u_g} \rangle}$ which informs about the relative drainage of liquid through the moving foam. The absence of relative drainage,  
 ${\langle \overline{u_l} \rangle}/{\langle \overline{u_g} \rangle}=1$, implies that $\alpha_l=\Phi_l$, while drainage in the direction of the flow, ${\langle \overline{u_l} \rangle}/{\langle \overline{u_g} \rangle}>1$, entrains a injected liquid fraction higher than the volume liquid fraction $\alpha_l>\Phi_l$.

\section{\label{sec:Formation} Microfoam formation at low liquid content}

\subsection{Bubbling regimes}
To study bubble formation and the accompanying bubble topology in the channel we vary the gas pressure at constant liquid flow rate. See fig. \ref{fig:regimes} for examples of the observations in the 250 $\mu$m high channel, near the 
orifice at the channel entrance for a fixed flow rate of $Q_l=167$ $\mu$l/min.  Above a certain threshold in gas pressure $P_g$ bubbles form in the channel. The bubble volume grows when increasing $P_g$, inducing several regimes of bubble formation and flow. 
  
We observe a minimum pressure $P_c$ for which bubbles form. For lower pressures the gas-liquid interface does not enter the orifice. At this liquid flow rate $P_c=0.9 \pm 0.15$ kPa (Fig. \ref{fig:regimes}a).
This effect is probably due to the capillary pressure. 
For a curved interface in the orifice considering the limit of bubble detachment, the Laplace pressure of a wetting interface is 
\begin{equation}
P_\sigma=\sigma \left(\frac{1}{r_1} + \frac{1}{r_2}\right)=1.1 \pm 0.2 \quad \mathrm{kPa},
\label{eq:Pcapillaire}
\end{equation}
 where $r_1=h/2$ and $r_2=w_{or}/2$ are the principal radii of curvature,  of the same order as $P_c$.

Above the initial pressure $P_c$,  a gas thread is forced into the orifice and fills a bubble after the orifice. This thread pinches off and releases the bubble. After break-up, the gas-liquid interface retracts to its initial position, as reported in \cite{Garstecki2005}, returning completely into the upstream part (dripping flow (i), Fig. \ref{fig:regimes}b).  At higher $P_g$, there is a coexistence, probably indicating a  first-order transition, with a second mechanism, where the interface 
remains in the orifice instead of retracting after bubble release. For given $P_g$ and 
$Q_l$, both mechanisms give different volumes $V_1$, $V_2$. This results in a flow of 
period $T_1+T_2$ \cite{Garstecki2004}, with bidisperse bubbles (ii, Fig. \ref{fig:regimes}b).

Further increasing $P_g$,  we only observe the second pinch-off mechanism, always resulting in a monodisperse foam. Three possible 
structures appear, according to the flow rate: bubbly flow (iii), alternate  foam with two rows (iv), or bamboo foam with one row only (v). No multiple-period or chaotic bubbling is observed. This suggests the absence of inertial non-linearities during the retraction of the gas-liquid interface \cite{Garstecki2005b}.  

For much higher $P_g$ the gas thread stops breaking up and a stratified liquid-gas flow is observed.

\subsection{Microbubble volume}
\begin{figure}[htbp]
{\includegraphics[scale=0.666]{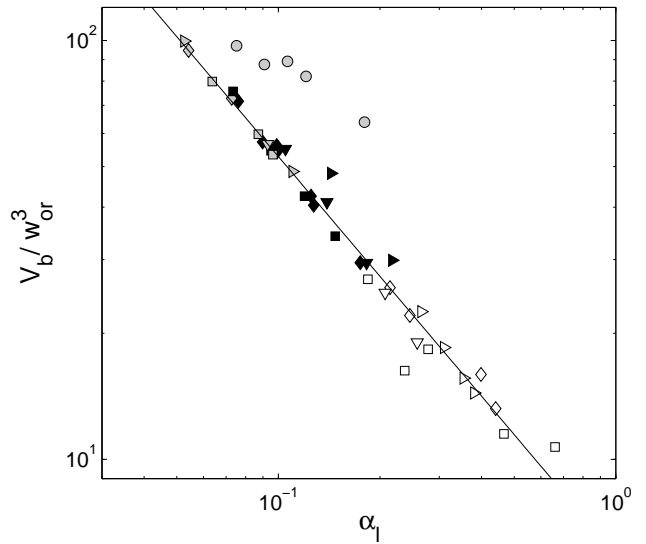}}
        \caption{The bubble volume $V_b$, in units of $w_{or}^3=1$ nl, depends only on the injected liquid fraction $\alpha_l$. The solid line is a linear fit to all data except the lowest $Q_l$: $\log (V/w_{or}^3) = (-0.95 \pm0.02) \log \alpha_l + (1.78\pm0.05)$.
}
        \label{fig:VbQgQl}
\end{figure}

The bubble volume for the foam regimes (iii, iv and v)
correlates well with the fluid fraction: 
\begin{equation}
\frac{V_b}{w_{or}^3} \sim \alpha_l^{-0.95\pm0.02}, 
\end{equation}
see Fig. \ref{fig:VbQgQl},  except for the 
lowest $Q_l$. At very low liquid content $\alpha_l \simeq Q_l/Q_g\ll 1$, bubble volumes are approximately proportional to $ (Q_l/Q_g)^{-1}$.

This correlation is similar to the one observed in axisymmetric conditions by \cite{Ganan-Calvo2001,Ganan-Calvo2004} who measured $V_b/w_{or}^3 \sim (Q_l/Q_g)^{-1.11\pm0.02}$,  \emph{in the opposite case} of high liquid fraction with $Q_l/Q_g>5$, with separated bubbles, while here bubbles are in contact or in short spacing after formation. It differs from the $V_b\sim P_g/Q_l$ scaling observed by Garstecki et al.  \cite{Garstecki2004}: we will see below that  the gas flow rate and applied pressure are not proportional for a low liquid fraction microfoam in a channel. Moreover ref.  \cite{Garstecki2004} investigates high liquid fraction flows at channel aspect ratio of 0.04  (flatter channels) much lower than the present 0.3. Both points could explain the experimentally observed differences. 

In this geometry,  one or two rows of bubbles are observed. The transition from bubbly flow to a foam in which bubbles touch each other  is governed by the volume liquid fraction $\Phi_l$,  which is related to $\alpha_{l}$ through equation \ref{eq:massconservation}. 
Therefore increasing the number of bubble rows when the foam state appears would either require decreasing $V_b$ at given $\Phi_l$ or enlarge the space for newly formed bubbles. The former can be achieved for instance with an orifice aspect ratio $w_{or}/h$  closer to $1$  to restrict liquid flow; and the latter by a lower ratio of orifice to channel width $w_{or}/w$.

\begin{figure}[htbp]
{\includegraphics[scale=0.666]{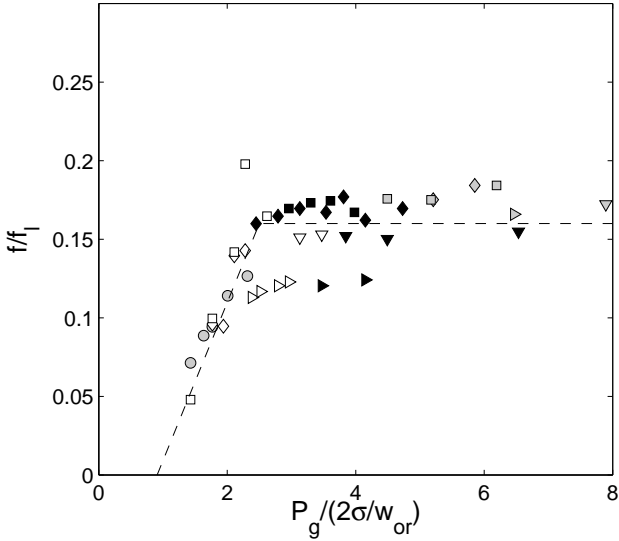}}
        \caption{The bubble formation frequency $f$ in units of $f_l=Q_l/w_{or}^3$ \textit{vs.} gas pressure $P_g$ in units of $2\sigma/w_{or}$. It shows a plateau for high $P_g$: the dotted line is a piecewise linear fit. Symbols correspond to different values for the liquid flow rate $Q_l$: (circle) 4, (diamond) 15, (square) 20, (right-pointing triangle) 30 and (down-pointing triangle) 40 $\mu$l min$^{-1}$, for bubbly flows (open symbols),  alternate foams (filled symbols), bamboo foams (gray symbols).
}
        \label{fig:Frequency}
\end{figure}

\subsection{Microbubble formation frequency}
The frequency $f$ of bubble formation (Fig. \ref{fig:Frequency}) first increases linearly then reaches a plateau for increasing $P_g$. The typical time and frequency linked to the liquid flow 
\begin{equation}
\tau_l=\frac{1}{f_l}=\frac{w_{or}^3}{Q_l}
\end{equation}
 can be used to define a non-dimensional frequency, known as the Strouhal number,  
\begin{equation}
St=\frac{f}{f_l}=\frac{f w_{or}^3}{Q_l}.
\end{equation}
After rescaling by $f_l$ all data collapse on a single curve where $St$ is a function of gas pressure only. Two regimes are observed: for low gas pressures in the case of bubbly flow we find $St\simeq 0.1 (P_g-P_c)/(2\sigma/w_{or})$, while for higher $P_g$, the Strouhal number saturates to a constant value of $St=0.16$. 


We infer that these two regimes are the consequence of two stages during the bubble formation:
\begin{enumerate}
\item Gas filling of the orifice:  At low $P_g$ the frequency varies like $f\sim f_l(P_g-P_c)\sim Q_l(P_g-P_c)$.
In other words, the period is proportional to a  characteristic time that varies as 
\begin{equation}
T\simeq\tau_{g}\sim[Q_l(P_g-P_c)]^{-1}.
\end{equation}
We interpret this time as the time necessary for the gas to fill the orifice, prior to break-up. It decreases with increasing $P_g-P_c$ since the gas pushes the fluid with a velocity increasing with pressure. $\tau_{g}$ also decreases for increasing $Q_l$ with the flow focusing confining more and more the available space for the gas thread. Note that the relation $f\sim Q_l P_g$ was proposed by \cite{Garstecki2004}, verified for varying $Q_l$ but with a constant $P_g$: here we also investigate the effect of gas pressure.
\item  Liquid mediated thread pinch-off: For high $P_g$ the bubble formation frequency is proportional to 
 $f\sim Q_l$. The period 
  only depends on 
  \begin{equation}
T\simeq \tau_l= \frac{w_{or}^3}{Q_l},
\end{equation}
 the time to pinch off the gaseous thread when the liquid flow is blocked by the bubble at the outlet \cite{Garstecki2005}.
\end{enumerate}

The bubbling period at low pressure is limited by the gas filling, while it is limited by the liquid driven thread contraction in  the high gas pressure regime.   Note that the transition from a $\tau_{g}$ to a $\tau_{l}$ dominated break-up frequency is accompanied by the regime change from bubbly flow to foam. In the foam state, the liquid flow restriction seems more efficient (see figure \ref{fig:regimes}b, iv and v). 

As a conclusion, there are two stages during formation: the first associated  with the filling of the orifice by the gas ($\tau_{g}$) and the second reflecting the pinch-off of the gas thread ($\tau_l$). They have different  gas pressure dependency ($\tau_g$ depends on gas pressure while $\tau_l$ does not), which creates a cross-over apparent in the bubbling period $1/f$ that depends on $\tau_{g}$ and $\tau_l$. 

\section{\label{sec:Flow} Foam flow}

We now turn to the flow of a foam in the microchannel after formation. When we measure the average gas flow rate as a function of  the applied gas pressure drop $P_g$,  we observe a highly non-linear response (see Fig. \ref{fig:qgpg}).  We find: a threshold (section \ref{sec:Pressurethreshold}), a non-linear slope (section \ref{sec:Non-linear flow-rate to pressure dependence}) and a discontinuity upon the transition from an alternate to a bamboo foam (section \ref{sec:Discontinuities in the flow-rate}).  The pulsation  of the flow rate at the  bubble formation frequency is finally examined (section \ref{sec:Flow pulsation at bubbling frequency}).
 
 \subsection{Pressure threshold}
 \label{sec:Pressurethreshold}
\begin{figure}[htbp]
\setlength\unitlength{1 cm}
     \begin{picture}(8,8)(0.4,0)
    \put(0.02,0){\includegraphics[scale=0.666]{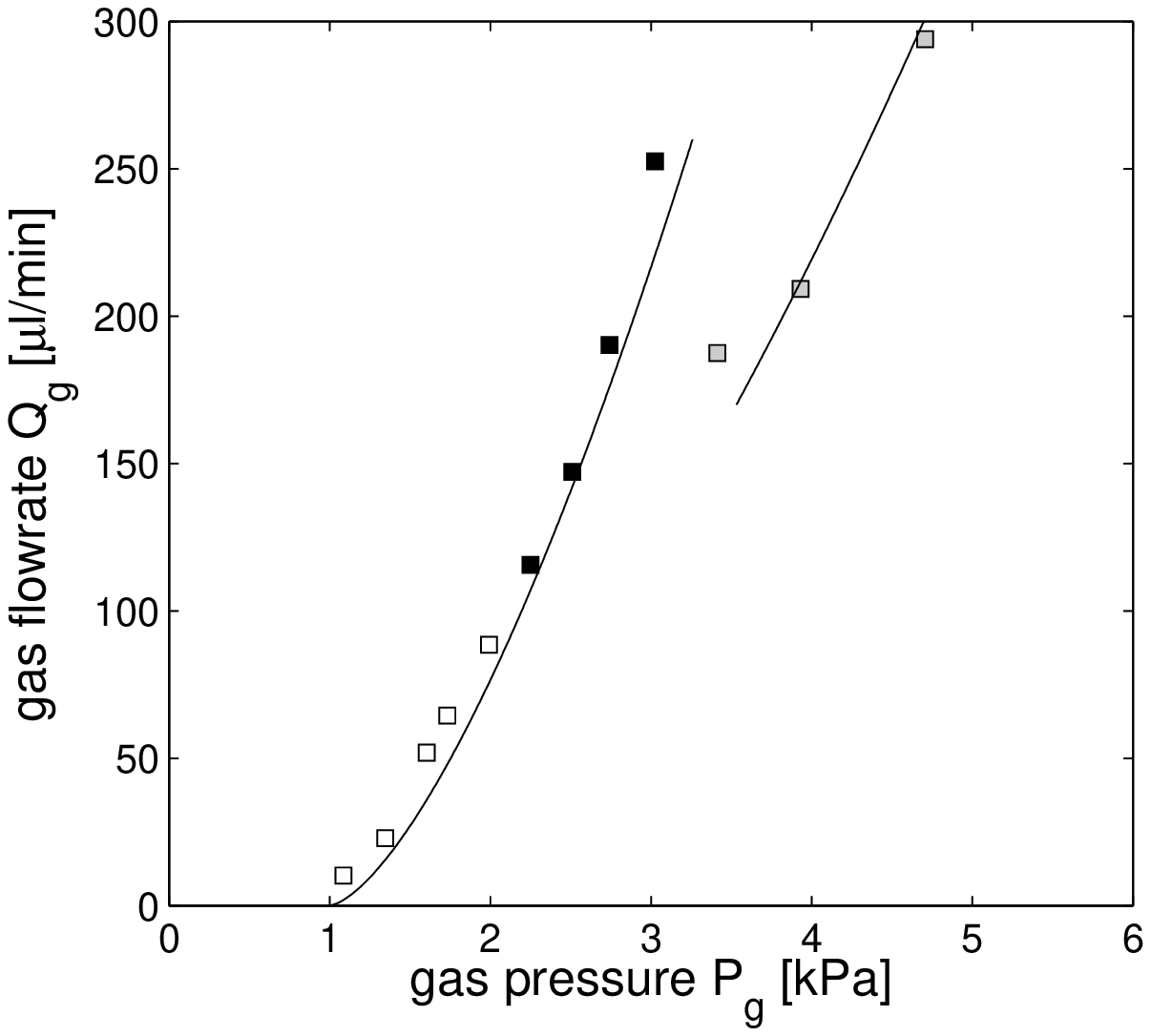}}
    \put(1.4,1){\includegraphics[scale=0.3]{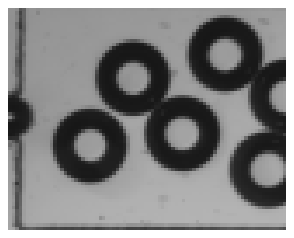}}
    \put(1.8,2.25){\includegraphics[scale=0.3]{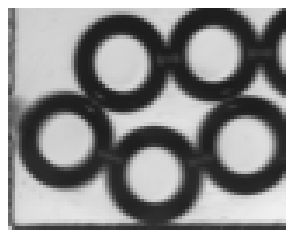}}
    \put(2.4,3.5){\includegraphics[scale=0.3]{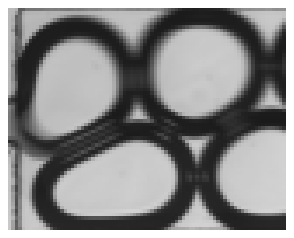}}
    \put(3.1,4.75){\includegraphics[scale=0.3]{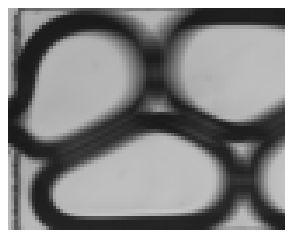}}
    \put(3.4,6){\includegraphics[scale=0.3]{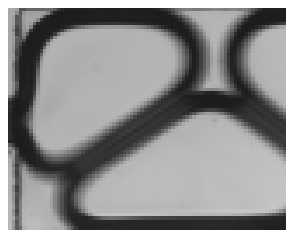}}
    \put(6.2,5){\includegraphics[scale=0.3]{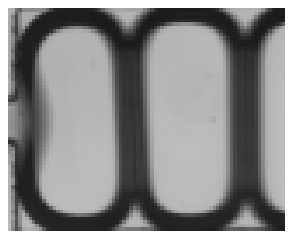}}
    \end{picture}
    \caption{\label{fig:qgpg} Gas flow rate $Q_g$ \textit{vs.} gas pressure $P_g$
showing a capillary threshold and discontinuity at the transition from alternate to bamboo foam  ($Q_l$ = 20 $\mu$l min$^{-1}$). 
}%
\end{figure}

Fig.  \ref{fig:qgpg} shows a threshold in 
pressure for the establishment of bubbly flow. It is found to be 1.0 $\pm$ 0.1 kPa, if this parameter is left 
free in the fit. It is compatible with the above explanation by a capillary 
effect (1.1 $\pm$ 0.2 kPa according to equation \ref{eq:Pcapillaire}) at the orifice. 

In presence of a second constriction (data not shown), we obtain for regimes (iii) and (iv)
the same result as in Fig. (\ref{fig:qgpg}), 
translated by about 0.45 kPa along the $P$-axis (compatible 
with the expected Laplace pressure necessary to overcome the second 
constriction, 0.6 kPa). This confirms that the threshold is induced by capillary effects. 
On the other hand, the slope originates from dissipative effects in 
the channel.  We thus write the total pressure drop as the sum of two contributions, 
\begin{equation}
P_g = P_c+\Delta P_{channel},	
\end{equation}
 where $P_c$ is the static orifice contribution and  $\Delta P_{channel}$ is due to dynamic dissipation in the channel. 

\begin{figure}[htbp]
{\includegraphics[scale=0.666]{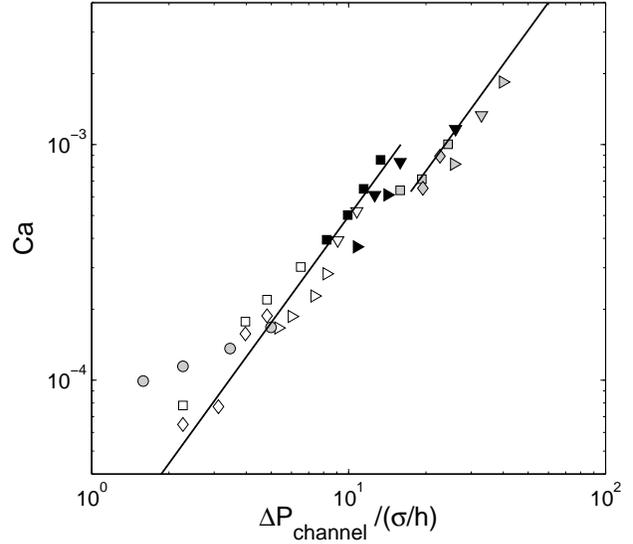}}
    \caption{\label{fig:PgCa} Graph using adimensional axes  $Ca=\mu Q_g/S\sigma$ versus $\Delta P_{channel}=P_g-P_c$. Data for different liquid flow rates are superimposed, $Q_l$ =  ($\diamond$) 4,  ($\Box$) 20, ($\triangleright$) 30
and ($\triangledown$) 40 $\mu$l min$^{-1}$.  The solid lines on both plots are fit to data on alternate and bamboo foams at $Q_l$ = 20 $\mu$l min$^{-1}$ with the power law $\Delta P_{channel}/(\sigma/h)=\beta \, Ca^{2/3}$,  with $\beta=1.7\times 10^{3}$ (alternate) and $\beta=2.5\times 10^{3}$ (bamboo).
}%
\end{figure}

\subsection{Non-linear flow-rate to pressure dependence}
\label{sec:Non-linear flow-rate to pressure dependence}
The flow rate is highly non-linear above the threshold, see fig \ref{fig:qgpg}.  The gas flow rate increases faster than a linear function of pressure for the alternate foam structure.

It can be interpreted by assuming that dissipation mainly  occurs in the liquid films, close to the walls \cite{Cantat2004}. Introducing the capillary number $Ca=\mu v/\sigma$ (of order $10^{-4}-10^{-3}$) containing the bubble velocity $v$ (estimated as $v\simeq Q_g/S$ in the dry foam state) and liquid viscosity $\mu$, the pressure drop writes
\begin{equation}
\Delta P_{channel}=\overline{\lambda}\frac{n L_{proj}}{S}\sigma Ca^{2/3},
\label{eq:Cantat}
\end{equation}
  with $n$ the total number of bubbles, $L_{proj}$ the projection on the cross section of the wetting perimeter per bubble, $S$ the cross section area, and $\overline{\lambda}$ a numerical constant \cite{Cantat2004}.  The effect of the orientation of the films relative to the foam movement is included in the $L_{proj}$ variable ($L_{proj}=L \cos\alpha$ for a film whose normal vector is slanted by an angle $\alpha$ with respect to the flow direction).  

In each regime, a fit of the pressure drop  $\Delta P_{channel}$ by $Ca^{2/3}$ gives a  correct agreement, see figure \ref{fig:PgCa}. Compared to a Newtonian flow whose drag pressure grows proportionally to the flow rate, the drag pressure grows with a lower exponent of the flow rate. It is an effect of the lubrication films between bubbles and walls that thicken when flow rates increase, a phenomenon described first in  \cite{Bretherton1961}, for  the motion of a single in a capillary tube. 

We measure the projected friction  length $L_{proj}$ as the projection of the  length  between vertices centers, for contacts with the walls parallel to the image plane, neglecting side walls. We deduce  for the numerical constant the value $\overline{\lambda}=22\pm5$.  The liquid coflow, varying over a decade, does not influence much the gas pressure drop in our geometry and the value of this constant (see  fig \ref{fig:PgCa}). 

We can compare the constant $\overline{\lambda}$ between our microchannel and the millimetric channels with a comparable aspect ratio (but without liquid coflow) studied by \cite{Cantat2004}, who found $\overline{\lambda}=38\pm04$.  We observe less friction: in the present experiments liquid is injected continuously, possibly explaining wetter foams and a lower value for $\overline{\lambda}$.   Since we see no visible change in the thickness of lubricating liquid films between gas and walls, which would change drag forces, we assume that liquid flows mostly in the corners \cite{Wong1995}. Thanks to  this corners the liquid flow is not as obstructed by  bubbles in the channel as with a cylindrical geometry \cite{Bretherton1961}. 
 
We find that the  measured foam liquid fractions is close to the injected fraction $\Phi_l \simeq \alpha_l$: the liquid fraction is here controlled by the injection parameters. Using  (\ref{eq:massconservation}) we obtain ${\langle \overline{u_l} \rangle}/{\langle \overline{u_g} \rangle}\simeq 1$: there is no substantial relative drainage of the liquid within these foams. From conservation of liquid mass in the corners, we deduce that an increase of the liquid flow rate with no variation in average velocity should lead to an increase of the diameter of the corner. 

\subsection{Discontinuities in the flow-rate} 
\label{sec:Discontinuities in the flow-rate}
The structure transition from an alternate to a bamboo foam induces a discontinuous decrease of the gas flow rate  (Fig. \ref{fig:qgpg}). 

This is a signature of the discrete character of the foam:  We find by image analysis that the transition to bamboo structure is associated with an increase in the bubble number $n$ of 15 \% (compactification). It is also associated with an increase of the projected length $L_{proj}$ of 50 \%. Both factors are consistent with the observed increase of $\Delta P_{channel}$ by  50 \%, considering equation \ref{eq:Cantat} that states that $\Delta P_{channel}$ is proportional to $n L_{proj}$. We conclude that the rearrangement of bubbles induces the discontinuity in the pressure drop. 

A consequence of this findings is that, in some ranges,  a given flow rate (here for $Q_g$ in between 170 and 260  $\mu$l min$^{-1}$) can lead to two possible  foam states, each one  associated with a different pressure drop.  
 
\subsection{Flow pulsation at bubbling frequency}
\label{sec:Flow pulsation at bubbling frequency}

\begin{figure}[htbp]
\setlength\unitlength{1 cm}
\begin{picture}(8,4.2)(-0.8,0)
	 \put(0,0){a)}
  	\put(3.7,0){b)}
	\put(.5,4){{\footnotesize x (mm) \quad 1 }}
	\put(-0.5,3){{\footnotesize t (s)}}
	\put(-0.3,.5){{\footnotesize 0.5}}
	\put(4.1,4){{\footnotesize 4}}
	\put(5.3,4){{\footnotesize 5}}
	\put(0.3,0){\includegraphics[scale=0.48]{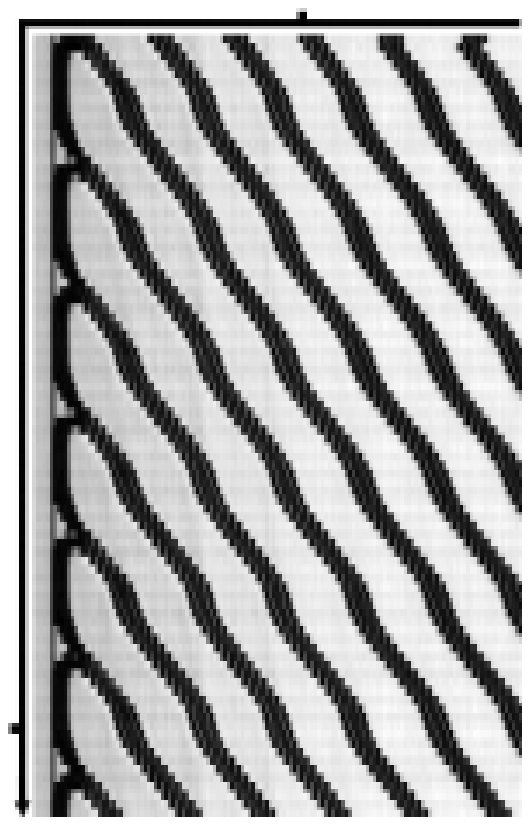}}
       \put(4,0){\includegraphics[scale=0.48]{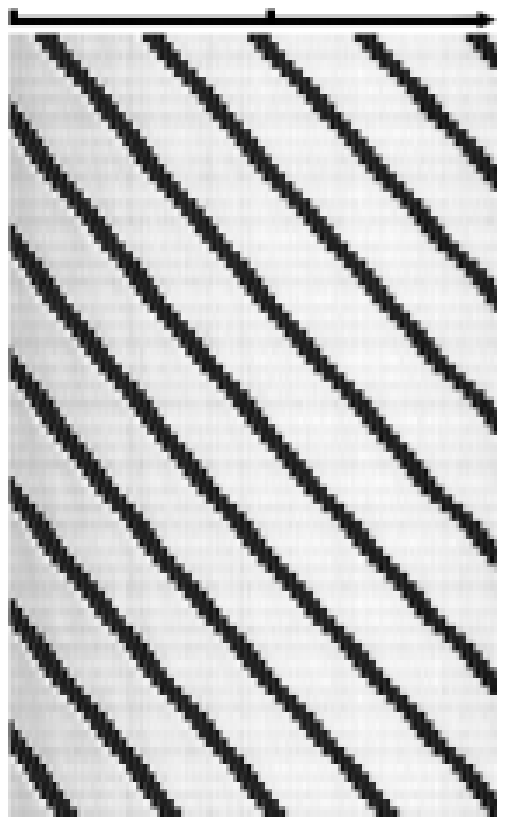}}
	\end{picture}
	\caption{\label{fig:timespace}
Space-time diagrams of the foam flow in the constriction sample:  (a) at the channel entrance; and (b) just before the exit constriction. The
vertical axis is the time, flowing downwards; the horizontal axis is
a (small zone of) the axis of the channel, with the foam flowing from
left to right, $Q_l=16.7$ $\mu$l min$^{-1}$ and $P=4.9$ kPa. Dark
pixels indicate a bubble edge. 
}
\end{figure}

 The above flow rate measurements are in fact {time-averaged} flows. The foam velocity at entrance oscillates at frequency $f$ (Fig. \ref{fig:timespace}a), between 3.1 and 8.3 cm/s.  However, this oscillation is damped over a few mm along the channel (data not shown). Very little oscillation is observed at the channel exit (Fig. \ref{fig:timespace}b).

Due to the periodic creation of bubbles, the pressure in the entrance orifice probably varies at frequency $f$. Each new bubble has to push the foam to create a place for itself; thus inducing the flow oscillation.  Friction, probably mainly at the channels walls,  homogenizes the flow at the micrometer scale. The forcing frequency is therefore gradually attenuated, allowing to consider the flow as steady further in the channel, as if pushed by a steady pressure. In view of the use of foams in channels with obstacles, this would prevent resonance effects with the frequency of passage of bubbles on the obstacles downstream.

\section{\label{sec:ultraflat} Ultraflat foam: distortion effects}

\begin{figure}[htbp]
	(a)
	\includegraphics[width=0.2\textwidth]{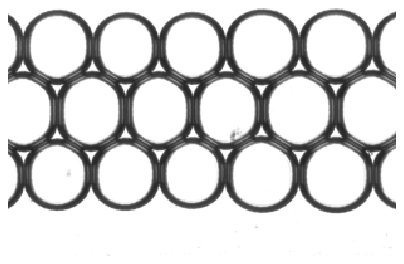}
	(b)
	\includegraphics[width=0.2\textwidth]{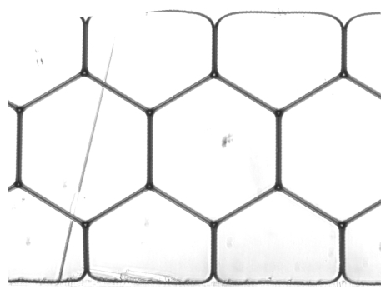}
	\\ (c)
	\includegraphics[width=0.2\textwidth]{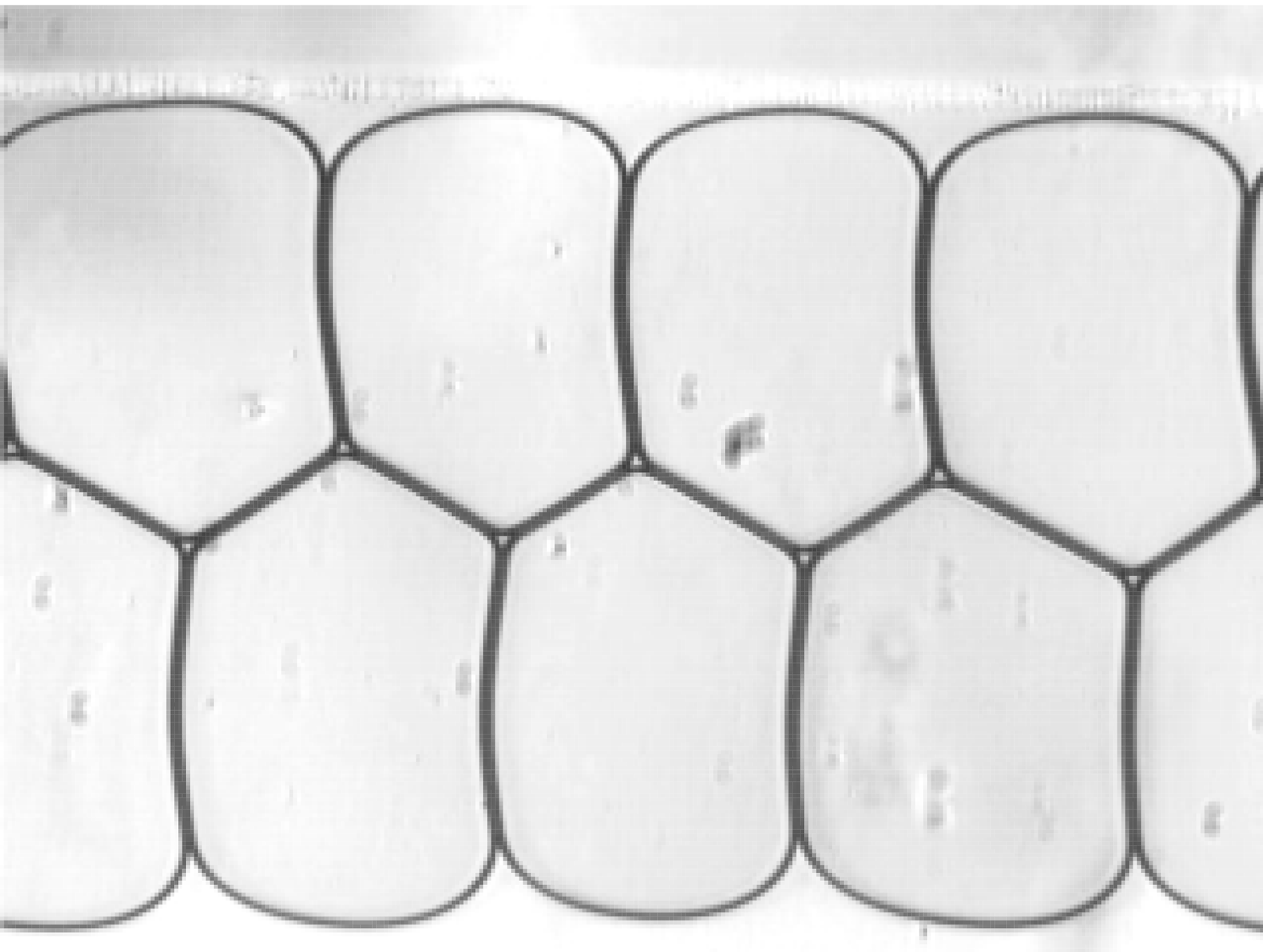}
	(d)
	\includegraphics[width=0.2\textwidth]{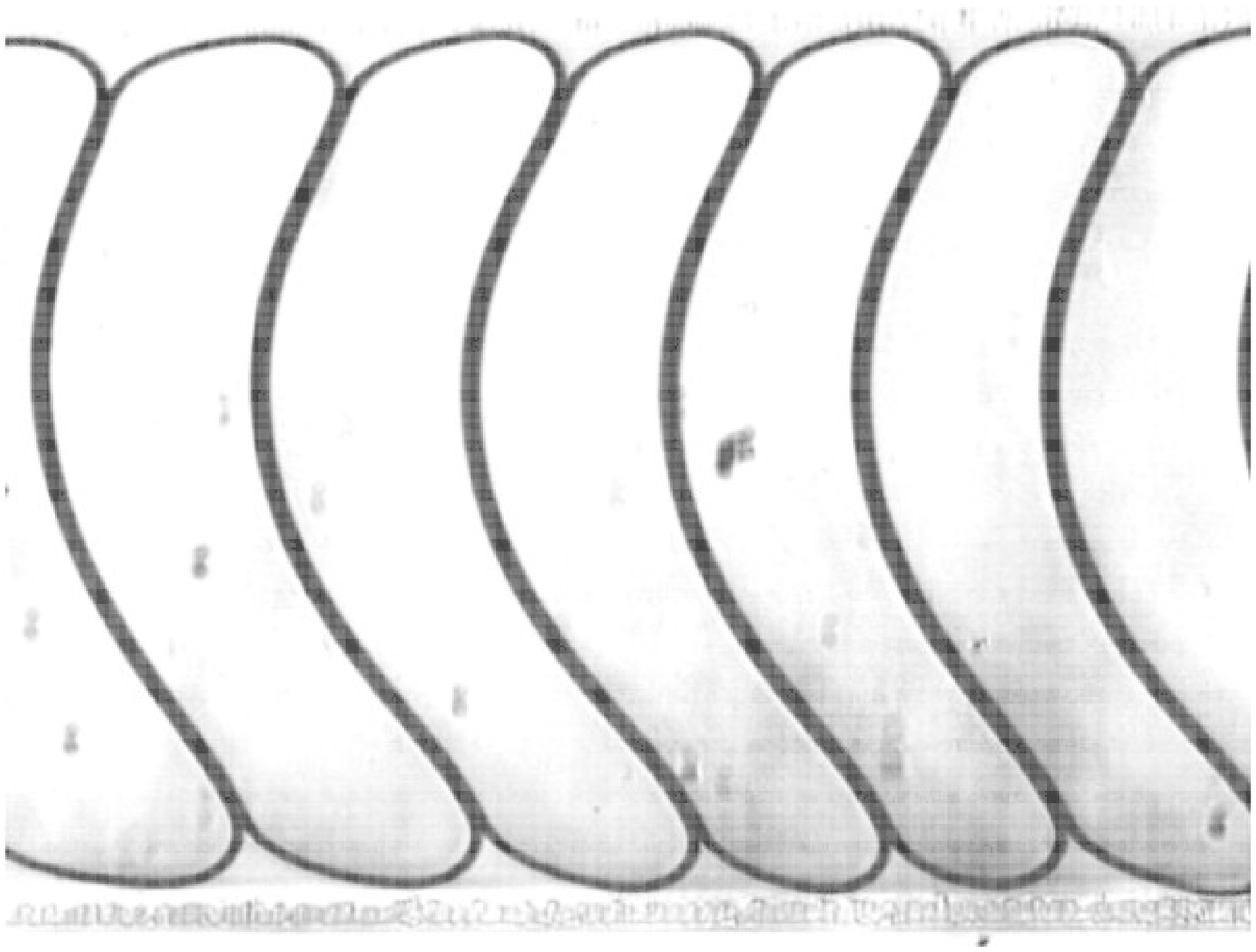}
	\caption{\label{fig:8micron} Ultraflat foams  
in the 8 $\mu$m high channel:
	 (a) flowing, wet and  (b) static, dry  3-rows foams;  (c) 2-rows and (d) 1-row flowing dry
boomerang foams. Flows from left to right. }
\end{figure}

In order to downscale even more flowing foams, we reduced the channel height to produce ultra-flat foams.  The ultra-flat channel presents a 30-fold decrease in height, to 8 $\mu$m, and a 18-fold decrease in aspect ratio for the channel section, to 0.02.
We continuously produce various foams in such a channel (Fig. 
\ref{fig:8micron}), including one with 3 bubble rows (Fig. 
\ref{fig:8micron}a). 

We can dry it, using the following batch method. We shut 
the liquid inlet
and pull the syringe at the gas inlet. As long as the 
underpressure is smaller than the
Laplace threshold (here 11 kPa), the bubbles are blocked by the 
orifice and only liquid flows out of the foam. This forced drainage 
yields hexagons with a standard deviation in the edge length of only $1.8$ \% (Fig. \ref{fig:8micron}b).
Since the apparent wall thickness on images (10 $\mu$m) is comparable to the height, 
the bubble walls are probably very curved (when looking at their  profile on a cross-section perpendicular to the image plane), and with {no 
 flat film between bubbles}, contrary to the previous  set-up. Thus the actual liquid fraction $\Phi_l$ of the 
central bubble row is probably smaller than, but close to, the apparent one (fraction of black pixels) $\Phi 
\approx\Phi_l^{image}\simeq 10^{-1}$. This contrasts with usual foams with 
larger aspect ratio, where the same picture of hexagons with 
straight walls and small vertices would correspond to much lower 
fluid fractions, $\Phi_l \simeq  10^{-2} <  \Phi_l^{image}$ \cite{Weaire1999}.

Foams flowing in this ultraflat channel undergo an unusual 
boomerang-like distortion, with films near the side walls pointing backwards (Fig. \ref{fig:8micron}c,d), the opposite of foams with no coflowing liquid \cite{Drenckhan2005}. We indeed expect a larger friction  at the center than on the edges, 
where bubble walls along the top and bottom plates are thin, than on 
the channel sides, where water accumulates. These edges are likely more entrained by a faster liquid flow than the centers (${\langle \overline{u_l} \rangle}/{\langle \overline{u_g} \rangle}> 1$, assuming that for small aspect ratios corner sections are less expandable). Within the reference frame moving at the liquid velocity,
the  bubbles move in the opposite direction, from right to left. 
The distortion of bubble shapes in this reference frame, with films near the walls pointing forward,  is then similar to the observations 
of millimetric bubbles pushed without any liquid flow \cite{Drenckhan2005}.

\section{\label{sec:conclusions} Conclusions}

We describe the formation and flow of a foam in a confined microchannel. The transition from bubbly flow to foam depends on $\Phi_l$ which is governed by the interplay between control parameters $P_g$ and $Q_l$ and the channel geometry. The frequency at which bubbles are formed behaves differently for bubbly flows and for foams. For foams the formation frequency only depends on the liquid flow rate controlling the speed at which the gas thread is pinched off. Foam flow and bubbly flow in microchannels are highly non-linear. The flow focusing orifice induces a threshold $P_c$ due to capillary effects in the flow-rate to pressure characteristic. The data for both bubbly and foam flow give good agreement to $P_g-P_c \sim Ca^{2/3}$. The prefactor in this relation depends on the dissipation in the channel related to the topology. 

Microfoams can be down-scaled to as small heights as  8 $\mu$m. The liquid fraction can be varied continuously over the complete range from the dry to the wet limit. We see an unusual deformation for foams flowing in this channel probably caused by relative drainage. 

The  interplay between geometrical parameters (channel aspect ratio, orifice aspect ratio and orifice to channel ratio)  merits more attention. 
It governs various effects as the transition from bubbly flow to foam,  the foam topology, and thereby dissipation in the channel, and the distortion of the foam cells. 
Comprehending this interplay will be a necessary step in the development of microfluidic applications in which the straight channel section will be replaced by more complex geometries allowing operations like mixing, separation, breaking and coalescence of bubbles.  
\section*{Acknowledgments}

We would like to thank  W. Drenckhan for stimulating discussions, and T. Podgorski  for his help on microchannel production. 


\begin{thebibliography}{13}

\bibitem{Link2004}
D.~Link, S.~Anna, D.~Weitz, H.A. Stone, Phys. Rev. Lett. \textbf{92}, 054503
  (2004)

\bibitem{Cubaud2004}
T.~Cubaud, C.M. Ho, Phys. Fluids \textbf{16}, 4575 (2004)

\bibitem{Garstecki2004}
P.~Garstecki, I.~Gitlin, W.~DiLuzio, G.~Whitesides, Appl. Phys. Lett.
  \textbf{85}, 2649 (2004)

\bibitem{Garstecki2005}
P.~Garstecki, H.~Stone, G.~Whitesides, Phys. Rev. Lett. \textbf{94}, 164501
  (2005)

\bibitem{Drenckhan2005}
W.~Drenckhan, S.~Cox, G.~Delaney, D.W. H.~Holste, N.~Kern, Colloids and
  surfaces A: Physicochem. Eng. Aspects \textbf{263}, 52 (2005)

\bibitem{Garstecki2005b}
P.~Garstecki, M.J. Fuerstman, G.M. Whitesides, Phys. Rev. Lett. \textbf{94},
  234502 (2005)

\bibitem{Weaire1999}
D.~Weaire, S.~Hutzler, \emph{The physics of foams} (Oxford University Press,
  1999)

\bibitem{Dou2002}
Y.H. Dou, N.~Bao, J.J. Xu, H.Y. Chen, Electrophoresis \textbf{23}, 3558Ð3566
  (2002)

\bibitem{Ganan-Calvo2001}
A.~Ganan-Calvo, J.~Gordilloa, Phys. Rev. Lett. \textbf{87}, 274501 (2001)

\bibitem{Ganan-Calvo2004}
A.~Ganan-Calvo, Phys. Rev. E \textbf{69}, 027301 (2004)

\bibitem{Cantat2004}
I.~Cantat, N.~Kern, R.~Delannay, Europhys. Lett. \textbf{65}, 726 (2004)

\bibitem{Bretherton1961}
F.P. Bretherton, J. Fluid. Mech. \textbf{10}, 166 (1961)

\bibitem{Wong1995}
H.~Wong, C.J. Radke, S.~Morris, J. Fluid. Mech. \textbf{292}, 71 (1995)

\end{thebibliography}

\end{document}